\documentclass[hidelinks,aps,pra,twocolumn,superscriptaddress,10pt]{revtex4-2} 

\usepackage{amsmath,amssymb}
\usepackage{mathrsfs}
\usepackage{epstopdf}
\usepackage{graphicx}
\usepackage{bm,color,bbm}
\usepackage{natbib}
\usepackage{hyperref}
\usepackage{caption}
\usepackage{fancyhdr}
\newcommand{\nn}{\nonumber}

\newtheorem{theorem}{Theorem}

\newtheorem{definition}{Definition}

\begin{document}
\fancyhead[R]{\ifnum\value{page}<2\relax\else\thepage\fi}


\title{Negativity vs. Purity and Entropy in Witnessing Entanglement}


\author{James Schneeloch}
\email{james.schneeloch.1@afrl.af.mil}
\affiliation{Air Force Research Laboratory, Information Directorate, Rome, New York, 13441, USA}

\author{Christopher C. Tison}
\affiliation{Air Force Research Laboratory, Information Directorate, Rome, New York, 13441, USA}

\author{H Shelton Jacinto}
\affiliation{Air Force Research Laboratory, Information Directorate, Rome, New York, 13441, USA}

\author{Paul M. Alsing}
\affiliation{Air Force Research Laboratory, Information Directorate, Rome, New York, 13441, USA}


\date{\today}

\begin{abstract}
In this work, we show that while all measures of mixedness may be used to witness entanglement, all such entangled states must have a negative partial transpose (NPT). Though computing the negativity of the partial transpose scales well at high dimension, it relies on knowing the complete quantum state, which does not.  To address this, we compare different measures of mixedness over uniform ensembles of joint quantum states at varying dimension to gauge their relative success in witnessing entanglement. In doing so, we find that comparing joint and marginal purities is overwhelmingly more successful at high dimension at identifying entanglement than comparing joint and marginal von Neumann entropies, in spite of requiring fewer resources. We conclude by showing how our results impact the fundamental relationship between correlation and entanglement and related witnesses.
\end{abstract}


\maketitle

\thispagestyle{fancy}
\section{Introduction}\label{Introduction}
Quantum entanglement is the principal resource consumed in many applications of quantum information such as quantum computing, communication, and enhanced quantum metrology. Understanding its fundamental nature goes hand in hand with developing adequate techniques to fully characterize it in the exceptionally high-dimensional systems being employed today, such as: quantum computation on 127-qubit states \cite{chow2021ibm}, boson samplers in $10^{30}$-dimensional state spaces \cite{Zhongeabe8770} or in pairs of particles entangled in high-dimensional degrees of freedom \cite{schneeloch2019record}. 

In this article, we begin by examining how measures of mixedness are used to witness entanglement, and provide a short proof demonstrating how all states whose entanglement is witnessed this way must have a negative partial transpose (NPT). The Peres Separability Criterion \cite{Peres1996} proves that all states with a negative partial transpose (known as NPT states) are entangled. This negativity is an entanglement witness that captures all entangled states that might otherwise be witnessed by comparing measures of mixedness. 

Because some measures of mixedness are more straightforward to obtain experimentally than the negativity, we compare how well two popular measures of mixedness demonstrate entanglement over large random ensembles of quantum states. In this work, we compare the effectiveness of demonstrating entanglement through comparisons of joint and marginal von Neumann entropy ($S_{1}(\hat{\rho}_{AB})$ and $S_{1}(\hat{\rho}_{A})$, respectively) to comparisons of the joint and marginal purity given as $\text{Tr}[\hat{\rho}_{AB}^{2}]$ and $\text{Tr}[\hat{\rho}_{A}^{2}]$, respectively. To facilitate a simpler side-by-side comparison, we use the negative logarithm of the purity given as the quantum collision entropy $S_{2}(\hat{\rho})$ instead of the purity itself. Note that the subscript $1$ or $2$ refer to different orders of $\alpha$ within the family of quantum Renyi entropies, which we will discuss in more detail later in the paper. In addition, the state purity can be used to bound the value of the von Neumann entropy (discussed in Section~\ref{bounds}), which is a more valuable measure in quantum information.

\section{Foundation: Entanglement from Mixedness and Majorization}\label{foundation}
In classical probability, joint distributions are never less mixed than the marginal distributions obtained from them \footnote{In the limit that the joint and marginal probability distributions are equally mixed, the joint distribution is a diagonal distribution with diagonal elements equal to the marginal probabilities. No joint distribution is less mixed because already single elements correspond to the marginal probabilities. All other joint distributions whose sums give the marginal distributions are necessarily more mixed because they can be obtained by mixing different permutations of the joint distribution.}. In the language of Shannon entropy, the joint entropy is never less than the marginal entropy; two random variables never take less information to communicate than one. However, this need not be the case when comparing the mixedness of joint and marginal quantum states.

To quantify the mixedness of quantum states, we measure the mixedness of the probability distribution generated by the eigenvalues of the density matrix. Given a probability distribution of $N$ outcomes $\{p_{i}\}_{i=1}^{N}$, we define the probability vector $\vec{p}$ as the $N$-dimensional vector whose components are the probabilities $\{p_{i}\}_{i=1}^{N}$. In addition, we provide the following definition for an arbitrary measure of mixedness:

\begin{definition}{}
A measure of mixedness is any continuous Schur-concave \cite{RobVarbConvexFunctions1973}\footnote{A Schur-concave function is a function $f:\mathbf{R}^{n}\rightarrow\mathbf{R}$ such that for any pair of vectors $\vec{u}$ and $\vec{v}$ in $\mathbf{R}^{n}$ in which $\vec{u}$ majorizes $\vec{v}$, it must follow that $f(\vec{u})\leq f(\vec{v})$.} function $\mathcal{F}$ of a probability vector $\vec{p}$ with minimum value zero for ``pure" probability distributions (in which one outcome contains all probability). The measure of mixedness for a quantum density matrix is of the probability vector of its eigenvalues.
\end{definition}
Such measures of mixedness $\mathcal{F}$ are also maximum for the uniform distribution, and monotonically increase under any mixing operations that replace elements of $\vec{p}$ with ones closer to the average value of the elements chosen. In particular, the value of $\mathcal{F}$ must increase for any distribution where unequal elements are re-distributed to bring them closer to their arithmetic mean (known as Robin-Hood operations  \footnote{Robin Hood mixing operations take two unequal elements of a probability distribution and bring them closer to their arithmetic mean, by adding to the smaller element and subtracting from the larger element an equal amount.}). All forms of entropy, including the von Neumann, Renyi, and Tsallis entropies are Schur-concave, and serve as measures of mixedness.

In examining measures of mixedness, there is a disconnect between showing that one distribution $\vec{q}$ is obtainable from another $\vec{p}$ through mixing operations, and that $\mathcal{F}(\vec{q})>\mathcal{F}(\vec{p})$ for some measure of mixedness. When (and only when) $\vec{q}$ can be obtained through a sequence of Robin-Hood operations on $\vec{p}$, we say that $\vec{p}$ \emph{majorizes} $\vec{q}$ \footnote{The definition of majorization is that $\vec{p}\succ \vec{q}$ if and only if for all $k$ from $1$ to the dimension of $\vec{p}$, the sum of the $k$ largest elements of $\vec{p}$ is greater than or equal to the corresponding sum of the $k$ largest elements of $\vec{q}$. Here, $\vec{p}$ and $\vec{q}$ have equal dimension.}, denoted by $\vec{p}\succ \vec{q}$. Alternatively, when the probability eigenvalues of density matrix $\hat{\rho}$ majorize the probability eigenvalues of density matrix $\hat{\sigma}$, then we say that $\hat{\rho}$ majorizes $\hat{\sigma}$ or that $\hat{\rho}\succ\hat{\sigma}$.  If $\vec{p}\succ \vec{q}$, then we know that the distribution $\vec{p}$ is more pure (less mixed) than $\vec{q}$ because there exists a series of mixing operations to obtain $\vec{q}$ from $\vec{p}$. That said, there are pairs of probability distributions where neither majorizes the other (here called incomparable), even though mixedness measures $\mathcal{F}$ are well-defined for both. This is because, for incomparable probability distributions represented by $\vec{p}$ and $\vec{q}$, one measure of mixedness $\mathcal{F}$ might show that $\vec{p}$ is more mixed than $\vec{q}$ via $\mathcal{F}(\vec{p})>\mathcal{F}(\vec{q})$, while another measure $\mathcal{G}$ might show that $\vec{p}$ is less mixed than $\vec{q}$ by $\mathcal{G}(\vec{p})<\mathcal{G}(\vec{q})$. However, when $\vec{p}$ and $\vec{q}$ are comparable (i.e., $\vec{p}\succ \vec{q}$ or $\vec{p}\prec \vec{q}$) then all measures of mixedness will agree on whether $\vec{p}$ is less mixed than $\vec{q}$.

Unlike classical probability distributions, quantum states are special because it is possible for the joint state of two parties $AB$ (given by the density matrix $\hat{\rho}_{AB}$) to be less mixed than the marginal state of either $A$ or $B$ (given by $\hat{\rho}_{A}$ and $\hat{\rho}_{B}$, respectively). For example, $AB$ can be in a pure quantum state $|\psi\rangle_{AB}$, such as a Bell state, while the reduced states of $A$ and $B$ are both maximally mixed. This can only happen, however, if the joint state is entangled \cite{PhysRevLett.86.5184}. In fact, it was proven in \cite{PhysRevLett.86.5184} that when $\hat{\rho}_{AB}$ is separable so that it has the form:
\begin{equation}\label{sepform}
\hat{\rho}_{AB}^{(sep)}\equiv \sum_{i} p_{i}(\hat{\rho}_{Ai}\otimes\hat{\rho}_{Bi}),
\end{equation}
then $\hat{\rho}_{AB}$ cannot be less mixed than either $\hat{\rho}_{A}$ or $\hat{\rho}_{B}$ because the probability eigenvalues of $\hat{\rho}_{AB}$ are \emph{majorized} by those of both $\hat{\rho}_{A}$ and $\hat{\rho}_{B}$. This is known as the majorization criterion of separability. Since all measures of mixedness cannot decrease under majorization, the majorization criterion of separability implies: for all measures of mixedness $\mathcal{F}$, separable states must satisfy the relation:
\begin{equation}\label{MixCrit}
\mathcal{F}(\hat{\rho}_{AB}^{(sep)})\geq\max\{\mathcal{F}(\hat{\rho}_{A}),\mathcal{F}(\hat{\rho}_{B})\},
\end{equation}
here called the mixedness criterion to distinguish it from majorization. However, the converse statement that all states satisfying the mixedness criterion satisfy the majorization criterion is demonstrably false. If a state satisfying the mixedness criterion for one measure of mixedness implied that the majorization criterion was satisfied, then it would also imply that the mixedness criterion is satisfied for all measures of mixedness. This is false because there exist states whose entanglement may be witnessed with one measure of mixedness, but not with another.

Comparing joint and marginal mixedness forms the basis of a broad class of entanglement witnesses. In addition to these entanglement criteria, there is another historical criterion relying on the form of the density matrix for separable states. Twenty-six years ago, Peres \cite{Peres1996} showed that where separable states \eqref{sepform} factor into products of states for each particle, and where the transpose of a density matrix is another valid density matrix, the \emph{partial} transpose of a separable state must also be a valid density matrix. Any state whose partial transpose yields a matrix with negative eigenvalues cannot be separable, and is therefore entangled. These entangled states are known as Negative-Partial-Transpose or NPT for short. Not all entangled states are NPT (though all 2-qubit entangled systems are \cite{Horodecki1996}), but it is a simple criterion to calculate from the density matrix, and functions based on these partial-transpose eigenvalues have been used as entanglement monotones (e.g., the negativity $\mathcal{N}(\hat{\rho})$ and log-negativity $E_{\mathcal{N}}(\hat{\rho})$).

While measures of mixedness are well-defined functions over all density matrices, it is possible (and common) for two density matrices to be incomparable with respect to each other (i.e., where neither density matrix majorizes the other). This incomparability suggests that there are states whose entanglement cannot be witnessed by comparing one measure of mixedness, but can by another, which motivates this study. Beyond this, we can also compare the set of states witnessed by violating the majorization criterion, to that of other separability criteria.

Before we show the details of our study comparing the relative effectiveness of different measures of mixedness at witnessing entanglement, we provide a short proof that the negativity of the partial transpose actually encompasses all comparisons of joint and marginal mixedness in their ability to witness entanglement. In particular, we prove that the set of states whose entanglement is witnessed by violating the majorization criterion (including those from violating \eqref{MixCrit}) is contained within the set of NPT states. In short, there are no entangled states violating the mixedness criterion \eqref{MixCrit} that are not also NPT, which is similarly easy to compute.

\section{The Supremacy of the negativity}\label{NegSuprem}
\begin{theorem}
Given a joint density matrix $\hat{\rho}_{AB}$,  if the mixedness criterion \eqref{MixCrit} is violated, then $\hat{\rho}_{AB}$ is NPT. Equivalently, the set of NPT states contains the set of states violating  the mixedness criterion \eqref{MixCrit}.
\end{theorem}

In 1998, the Horodeckis \cite{PhysRevLett.80.5239} showed that all states with a positive partial transpose are undistillable. Distilling entanglement involves taking copies of some partially entangled state, and then using Local Operations and Classical Communication (LOCC) to convert those states into fewer copies of maximally entangled states. Separable states are undistillable because LOCC operations cannot create entanglement, and separable states by definition have no entanglement to start with. Entangled states may or may not be distillable, where undistillable entangled states are known as bound-entangled. What the Horodeckis have shown is equivalent to the contrapositive statement \footnote{\label{MT}If a logical statement ``If $P$ then $Q$'' is true, then the contrapostive statement ``If not $Q$ then not $P$'' must also be true.} that all distillable states have a negative partial transpose (NPT). This is not the same thing as answering whether all NPT states are distillable, which remains an open question \cite{horodecki2020five}.

 In 2003, Tohya Hiroshima proved \cite{HiroshimaMajDistPRL2003} that if a joint state $\hat{\rho}_{AB}$ is undistillable, then it must satisfy the majorization criterion. This is equivalent to the contrapositive statement \cite{Note4} that all states that violate the majorization criterion (which includes those that violate the mixedness criterion \eqref{MixCrit}) must have distillable entanglement. Since we have previously established that all distillable states are NPT states, it follows that all states that violate the mixedness criterion in particular \eqref{MixCrit}, (and the majorization criterion in general) are also NPT states, thus proving Theorem 1.
\newline

By Theorem 1, we know there are no states that can violate the mixedness criterion \eqref{MixCrit} that won't also be NPT. The negativity will witness entanglement in at least all states whose entanglement can be witnessed by violating the mixedness criterion \eqref{MixCrit}. However, that does not mean that comparing measures of mixedness is obsolete. 

\begin{figure*}[t]
\centerline{\includegraphics[width=0.9\textwidth]{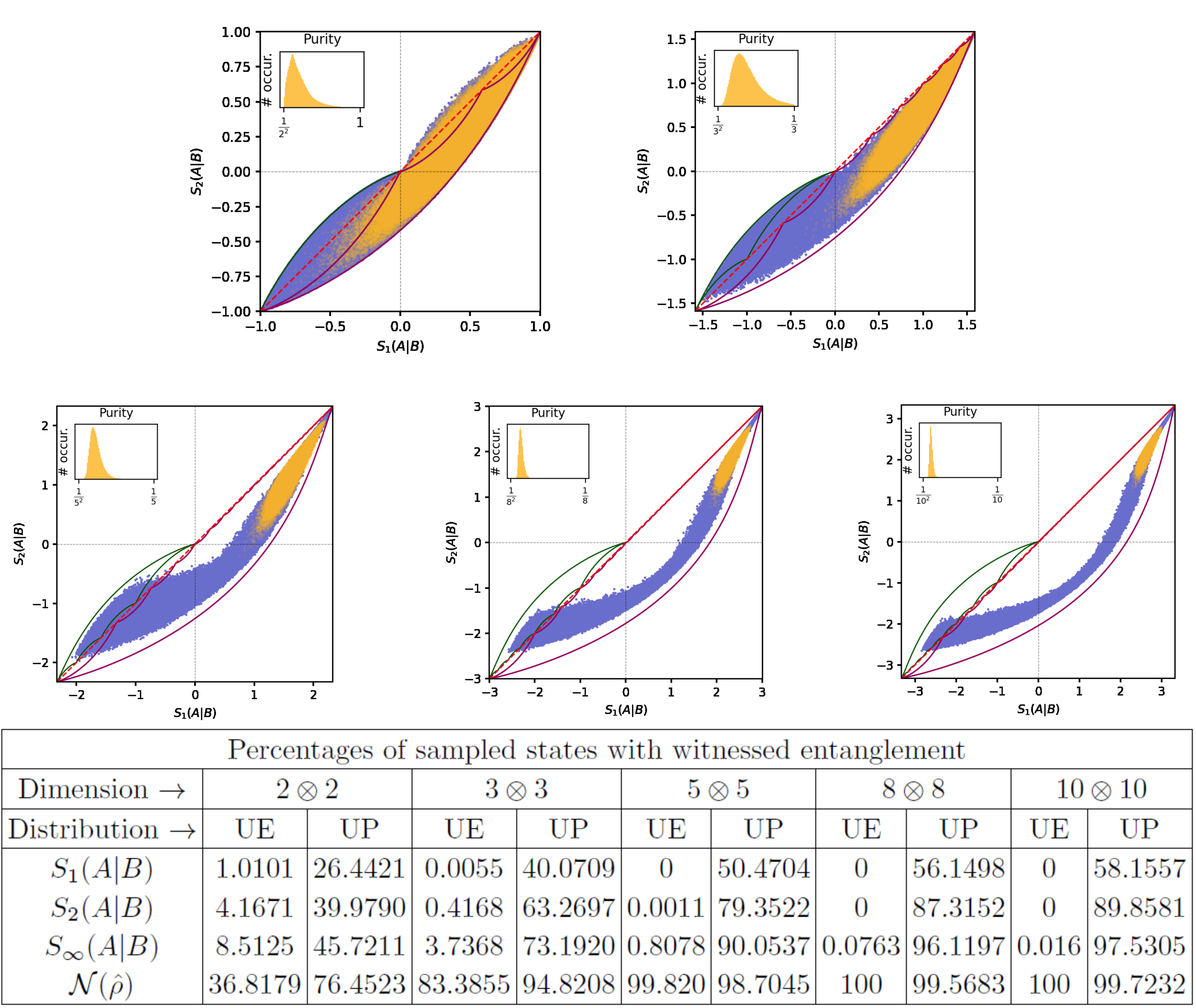}}
\caption{(top) Scatterplots of $S_{2}(A|B)$ vs  $S_{1}(A|B)$ and respective purity histograms for $10^{6}$ 2-quDit systems for $D=(2,3,5,8,10)$. The light orange scatterplots are from the fully uniform ensemble (abbreviated UE) while the blue scatterplots are from the ensemble uniform with respect to purity (abbreviated UP). The inset histograms are of the joint purity of the fully uniform ensemble). The red dotted line in each plot is where  $S_{1}(A|B)=S_{2}(A|B)$. The set of all $D\otimes D$ pure states is within the green serrated blade region in the lower left quadrant (or is only a single curve for $2\otimes 2$), while the set of all $D\otimes D$ states with maximally mixed marginals corresponds to the large magenta serrated blade spanning three quadrants of the plot. The regions enclosed between the two blades also correspond to valid density matrices.  (Bottom) This table gives the percentages of the total number of generated states whose entanglement was witnessed with the function in the first column.}
\end{figure*}

Although the negativity of the partial transpose is a computable entanglement witness from the density matrix, the difficulty in reconstructing a density matrix from experimental data would be intractable at high dimension due to the sheer number of elements that a density matrix may contain. Although tomography is not too challenging for a state made of one or two qubits, the number of elements to be determined increases exponentially with the number of qubits. For example, any quantum state in a Hilbert space of dimension $10^{30}$ can be expressed using no more than 100 qubits. In these regimes, no attempt at full state tomography would ever be made, but entanglement can be efficiently verified by obtaining a sufficiently high fidelity between the measured state and an ideal resource state \cite{guhne2009entanglement, PhysRevLett.92.087902}. What we explore here are strategies for efficiently characterizing the entanglement present in high dimensional systems where a target state is not given, but the state remains too large for tomography to be performed.

When full state tomography is not feasible, it is still theoretically possible to determine the eigenvalue spectrum of an $n$-dimensional density matrix $\hat{\rho}$ without determining the density matrix itself. If one can determine all of the $n$-th order moments of $\hat{\rho}$ given as the trace $\text{Tr}[\hat{\rho}^{n}]$ directly from experimental data, then the eigenvalue spectrum can be obtained by solving the set of $n$ eigenvalue equations. Though there exist multiple strategies for determining the second-order moment \cite{PhysRevLett.88.217901,brun2004measuring,brydges2019probing}, determining the $n$-th order moment requires measurements that interfere $n$ identical copies of the quantum state, which makes determining the full eigenvalue spectrum unfeasible at high dimension as well. There also exist strategies for estimating the negativity without having to determine the density matrix. These require determning at least the third-order moment of the partial transposed density matrix, which may be obtained experimentally through random or collective measurements \cite{yu2021optimal, Elben2020EntLocRandMeas, ZhouSingleCompEntNeg2020, GrayMachineAsstEntMeasPRL2018}. However, the \emph{second}-order moment of the density matrix is a valid measure of mixedness in its own right, known as the purity. In what follows, we show compelling evidence that comparing the joint and marginal purity witnesses entanglement more often than comparing the joint and marginal von Neumann entropy even though the latter is more difficult to determine experimentally. Moreover, we demonstrate how states for which comparing von Neumann entropies is more sucessfull are exceptionally rare in a uniform ensemble of density matrices.

\section{Quantum state purity \\vs von Neumann entropy \\in entanglement witnessing}\label{PurVsVNent}
In this section, we examine measures of mixedness based on the second-order moment of the density matrix (i.e., $\text{Tr}[\hat{\rho}^{2}]$), in comparison to the von Neumann entropy given as $-\text{Tr}[\hat{\rho}\log(\hat{\rho})]$. In particular, we show how comparing the joint and marginal state purities is almost always more successful at witnessing entanglement than comparing joint and marginal von Neumann entropies, even though fewer resources are required to determine the state purity. While the von Neumann entropy requires knowing the complete eigenvalue spectrum of the density matrix, state purities can be measured directly by interfering two identical copies of the system in an experiment \cite{PhysRevLett.88.217901,brun2004measuring}. To facilitate a side-by-side comparison of von Neumann entropy and state purity at witnessing entanglement, we consider comparing the Renyi entropies of order $\alpha$ (given by $S_{\alpha}$) without loss of generality:
\begin{align}
S_{\alpha}(A)&=S_{\alpha}(\hat{\rho}_{A})\equiv\frac{1}{1-\alpha}\log\Big(\text{Tr}[\hat{\rho}_{A}^{\alpha}]\Big),\\
\lim_{\alpha\rightarrow 1}S_{\alpha}(A)&=S_{1}(A)\equiv-\text{Tr}[\hat{\rho}_{A}\log(\hat{\rho}_{A})],\\
\lim_{\alpha\rightarrow 2}S_{\alpha}(A)&=S_{2}(A)=-\log\Big(\text{Tr}[\hat{\rho}_{A}^{2}]\Big),\\
\lim_{\alpha\rightarrow \infty}S_{\alpha}(A)&=S_{\infty}(A)=-\log\Big(\max_{i}\{\lambda_{i}\}\Big),\\
S_{\alpha}(A|B)&=S_{\alpha}(AB)-S_{\alpha}(B).
\end{align}
Here we see that $S_{1}(A)$ is the von Neumann entropy of system $A$, and $S_{2}(A)$ is a monotonically decreasing function of the purity $\text{Tr}[\hat{\rho}_{A}^{2}]$, known as the quantum collision entropy. In addition, $S_{\infty}(A)$ is known as the quantum min entropy, whose utility in entanglement witnessing is illustrated in Sec.~\ref{sidenote}. We define the Renyi conditional entropy $S_{\alpha}(A|B)$ for convenience. Whenever $S_{\alpha}(A|B)$ is negative, $S_{\alpha}(AB)$ is less than $S_{\alpha}(B)$, which witnesses entanglement by violating the mixedness criterion \eqref{MixCrit}. Here and throughout this paper, all logarithms are base two, since we measure entropy in bits.

\subsection{Monte Carlo simulations of random density matrices}\label{MCSim}
In order to compare the effectiveness of comparing von Neumann entropies to comparing state purities as witnesses of entanglement, we performed Monte-Carlo simulations on 1 million 2-quDit systems, for $D=\{2,3,5,8,10\}$. In other words, we randomly generated these 2-quDit systems and calculated their joint and marginal von Neumann entropies and purities to see what fraction of states generated had their entanglement witnessed by each measure of mixedness. For each dimension, we generated two uniform ensembles of density matrices. The first was completely uniform over the simplex of eigenvalues (discussed in the following paragraph), while the second ensemble was uniform over the simplex for each value of purity, but with the value of purity distributed uniformly as well. In Fig.~1 the orange scatterplots give the fully uniform ensemble (abbreviated as UE), while the blue scatterplots give the uniform purity ensemble (abbreviated as UP). The reason for generating the second distribution is because the fully uniform ensemble of density matrices produces nearly pure, nearly maximally mixed, and highly entangled states with negligible probability at high dimension, as discussed later in this section.

\subsubsection{Step 1: Generating the eigenspectrum}\label{eigenspectrum}
Generating a fair sampling of random density matrices is a two-step process discussed in \cite{PhysRevA.58.883, MezzadriRandomUnitary}. First, the eigenvalue spectrum of the density matrix is generated from a uniform distribution of probability vectors. This works because all probability vectors of dimension $N$ represent valid eigenvalue spectra for density matrices of dimension $N$ and vice versa. The uniform distribution of probability vectors is defined as follows. The set of probability vectors $\vec{p}$ of dimension $N$ forms a hyperplane of dimension $N-1$ due to the constraint equation that the sum of all components of $\vec{p}$ add to unity. This hyperplane is further bounded into a regular $(N-1)$-dimensional simplex by the constraints that each component of $\vec{p}$ be non-negative. In Fig.~3, we have a diagram of the uniform distribution of eigenvalue vectors for $N=3$. As a flat surface in $N$-dimensional space, the uniform distribution of probability vectors is uniform on this surface. In Fig.~1 the orange scatterplots and histograms refer to states generated from this uniform ensemble.

It may seem that great pains are taken to generate this particular distribution of eigenvalue spectra when one could otherwise simply generate random numbers between zero and one for each eigenvalue and renormalize to set the sum equal to one. However, such a process is overwhelmingly weighted in favor of the maximally mixed state at high dimension, due to the law of large numbers (see Appendix \ref{naive} for details).

When studying the effectiveness of various entanglement witnesses, it is important to cover all possible values that these witnesses might take with a large enough number of randomly generated states. Pure states of two or more parties that are anything other than an uncorrelated product of component party states are entangled. Nearly-pure states with slight but sufficient correlations are also demonstrably entangled (see equation \eqref{EntBoundCorrRel1} for illustration). For increasing levels of mixedness, there are fewer entangled states consistent with that amount of mixedness \cite{PhysRevA.58.883}. However, the uniform distribution of probability vectors has only a small fraction of its total hypervolume in close proximity to a pure state. Indeed, if we take the fraction of ``nearly-pure'' states to be the fraction of states that have a maximum probability component of at least $1/2$, that fraction of total probability vectors that are nearly pure would be $N\times2^{(1-N)}$, which decreases exponentially toward zero for large $N$. Alternatively, if we take the fraction of ``nearly maximally mixed'' states to be those with a purity $\mathcal{P}$ between $1/N$ and $1/(N-1)$, one can show the fraction of nearly maximally mixed states decreases even faster at high dimension (see discussion in Appendix \ref{nearlypure} for details and histograms in Fig.~1 for examples).

In order to better cover the full range of values that the quantum entropy can take, we created a second ensemble of probability vectors which, for a fixed value of purity, is otherwise uniform on the probability simplex. If we constrain the $N$-dimensional probability vectors on the simplex to also have a constant purity $\mathcal{P}$, the result is the intersection of a sphere of radius $\sqrt{\mathcal{P}-1/N}$ centered at the maximally mixed state (as illustrated in Fig.~3) intersecting the simplex. These uniform spherical slices of the probability simplex are far from straightforward to generate, but are described in \cite{Alsing2022}. With these uniform spherical slices, we generate the uniform-purity ensemble by generating a uniform random number for the purity, and using it as a seed to generate a random probability vector on the spherical slice corresponding to that purity. In Fig.~1, the blue scatterplots refer to states generated from this uniform-purity ensemble of eigenvalue vectors. Once we generated the probability vector defining a random diagonal density matrix, the next and final step was to rotate it by a random unitary transformation to complete the random quantum state generation.

\subsubsection{Step two: Generating a random unitary transformation}\label{unitary}
Once we have both ensembles of randomly sampled diagonal density matrices, we rotate them by taking randomly selected unitary transformations whose distribution is uniform according to the Haar ensemble \cite{MezzadriRandomUnitary}. This is accomplished by taking a random matrix of normally distributed complex numbers of mean zero and variance unity, and then using Gram-Schmidt orthonormalization on the resulting matrix to obtain a unitary one. 
Unitary matrices generated this way are uniformly distributed with respect to the Haar measure, and their uniformity is well-illustrated by the following point: if one produces a distribution of unitary matrices via this method, and then rotates each matrix in that distribution by the same (but arbitrary) unitary transformation, the distribution overall will remain unchanged. This invariance is similar to how a cluster of points uniformly distributed on the sphere remains uniformly distributed on the sphere however it is rotated.

\subsection{The data set of randomly generated \\
density matrices}\label{dataSet}
With the algorithm to generate random density matrices described, we generated two classes of ensembles of density matrices. The first ensemble has a uniformly sampled set of eigenvalues, while the second class of ensembles is uniform with respect to purity in order to generate more highly entangled states and better explore the effectiveness of different entanglement witnesses. In the table at the bottom of Fig.~1 we show that the percentages of randomly generated states whose entanglement is witnessed by mixedness measures $S_{\alpha}(A|B)$ increase when moving from the uniform to the uniform-purity ensembles and dramatically so at higher dimension. Indeed, the probability that a state selected from the uniform purity ensemble is ``nearly pure'' is bounded below by $1/2$, while it decreases exponentially toward zero for the fully uniform ensemble.

In Fig.~1, we show scatter plots of the von Neumann conditional entropy $S_{1}(A|B)$ versus the second-order Renyi conditional entropy $S_{2}(A|B)$ obtained from the purity. When a state demonstrates entanglement by showing $S_{2}(A|B)<0$ , but not by $S_{1}(A|B)>0$, we say that the collision entropy has the advantage. In the alternate situation, we say the von Neumann entropy has the advantage. For two-qubit states, there do not appear to be any states for which the von Neumann entropy has the advantage. For the fully uniform and constant purity ensembles plotted here, it also appears that there are no states where the von Neumann entropy has the advantage, which implies that comparing purities will always be a more sensitive entanglement witness than comparing von Neumann entropies. However, this is not entirely the case.

Prior to developing a method of sampling uniformly at constant purity \cite{Alsing2022}, we had generated ensembles that covered a larger range of purities by taking the fully uniform ensemble, raising the (diagonal) density matrices to a given power based on the marginal dimension $D$ of the $D\otimes D$ states, and renormalizing. This new distribution of density matrices was highly non-uniform at constant purity, but covered a larger range of purity values to fill out the scatterplots. In these ensembles, we found for $3\otimes 3$ and $5\otimes 5$ states, that there do exist anomalous states for which the von Neumann entropy has the advantage. Upon examining these ensembles of $3\otimes 3$ and $5\otimes 5$ anomalous states, we found that they all have at least one thing in common. The joint and marginal density matrices for these states have approximately equal rank in that the $D$ largest joint eigenvalues of the $D\otimes D$ system contain almost all of the total probability. With this restriction, the joint purity is approximately bounded by the same range of values as the marginal purity. As one can see in the in-set plots of the histograms of joint purity in Fig.~1, the likelihood of generating these anomalous states appears to be vanishingly small for the fully uniform ensemble. 

As for why the uniform-\emph{purity} ensemble also produces no anomalous states, we can consider the likelihood of a high-dimensional state for a given value of purity also having a low rank. For a joint density matrix of dimension $N=D^{2}$, the fraction of joint states of rank no larger than $D$ is essentially zero because such states would reside on facets or edges at the boundary of the probability simplex, being an infinitesimal fraction of the total volume. Where the anomalous states generated have only approximately low rank  (with otherwise many, but very small nonzero probabilities), the probability of generating states that are very close to these boundaries is still correspondingly small. This fact remains true, even when sampling uniformly at a constant purity, because a uniform sample of the $N-1$ dimensional probability simplex at constant purity is still an $N-2$ dimensional piecewise manifold. The additional constraint of the joint density matrix having rank no larger than $D$ places the sample at a boundary of this piecewise manifold.

From this, we see that for nearly all states, comparing purities will be a more sensitive witness of entanglement than comparing von Neumann entropies. However, it is worth pointing out that the number of randomly generated states required to fill the state space to a given average density increases exponentially with dimension, making the subsequent scatterplots more diffuse.

As a concrete example of these anomalous states, we can consider a mixture of three orthogonal, $3\otimes 3$ partially entangled states:
\begin{equation}
\hat{\rho}=p_{1}|\psi_{1}\rangle\langle\psi_{1}| + p_{2}|\psi_{2}\rangle\langle\psi_{2}| + p_{3}|\psi_{3}\rangle\langle\psi_{3}|
\end{equation}
such that
\begin{subequations}
\begin{equation}
|\psi_{1}\rangle = \sqrt{\lambda_{1}}|0,0\rangle + \sqrt{\lambda_{2}}|1,1\rangle + \sqrt{\lambda_{3}}|2,2\rangle
\end{equation}
\begin{equation}
|\psi_{2}\rangle = \sqrt{\lambda_{1}}|1,0\rangle + \sqrt{\lambda_{2}}|2,1\rangle + \sqrt{\lambda_{3}}|0,2\rangle
\end{equation}
\begin{equation}
|\psi_{3}\rangle = \sqrt{\lambda_{1}}|2,0\rangle + \sqrt{\lambda_{2}}|0,1\rangle + \sqrt{\lambda_{3}}|1,2\rangle
\end{equation}
\end{subequations}
Where $|\psi_{1}\rangle$, $|\psi_{2}\rangle$, and $|\psi_{3}\rangle$ are all mutually orthogonal, the joint entropy $S_{\alpha}(AB)$ is purely determined by the probability vector $(p_{1},p_{2},p_{3})$. Where the set of Schmidt coefficients associated to the measurement outcomes of system $B$ is the same for $|\psi_{1}\rangle$, $|\psi_{2}\rangle$, and $|\psi_{3}\rangle$, the marginal entropy $S_{\alpha}(B)$ is determined purely by the probability vector $(\lambda_{1},\lambda_{2},\lambda_{3})$. Because we can choose $(p_{1},p_{2},p_{3})$ independently of $(\lambda_{1},\lambda_{2},\lambda_{3})$, it is straightforward to make an anomalous state where these two probability vectors are incomparable, and where the von Neumann entropy has the advantage at witnessing entanglement.

To give an idea of how large the scatterplots in Fig.~1 might be with an exhaustive set of density matrices, we have used upper and lower bounds for von Neumann entropy for a constant collision entropy (i.e., constant purity) to enclose neighborhoods associated to broad classes of quantum states in Fig.~1. For both the set of pure states (small green blade), and states with maximally mixed marginals (large purple blade), the conditional entropies are expressed  (up to a constant offset) as either marginal or joint entropies. Where every point inside either blade and in the gap between them (explained momentarily) corresponds to a valid density matrix, we see that either ensemble of density matrices does not cover the full spectrum of values that these entropies can take, demonstrating their relative rarity. Even starting from a uniform distribution of pure states, the distribution of marginal eigenvalue spectra from these pure states is heavily weighted against high entanglement, as discussed in Appendix \ref{MargEnt}.

In the scatterplots in Fig.~1, the region enclosed between the two blades also corresponds to valid density matrices, and can be understood in the following way. The operation of mixing a pure state with a maximally mixed state is a continuous transformation of the density matrix, which must ultimately transform every pure state into one with a maximally mixed marginal, but which remains a valid quantum state for every value of mixing. Since the end points of the blade in the scatter plots are both pure states and ones with maximally mixed marginals, any curve connecting those two points that starts within the neighborhood must pass through every point in the gap between the two blades. Thus, there is a valid quantum state for every point in the gap between these two blades.

\subsection{Side note: Increased sensitivity when using higher-order entropies}\label{sidenote}
Using higher-order moments of the density matrix may yield more sensitive entanglement witnesses than the purity, but at the expense of becoming progressively more difficult to obtain from experiment. In particular, the direct measurement of $\text{Tr}[\hat{\rho}^{n}]$ requires interfering $n$ copies of the state $\hat{\rho}$, which becomes intractable as $n$ grows large. Indeed, determining the eigenvalue spectrum of a thirty-qubit quantum state would require interfering over $10^{9}$ copies of the state.

That said, it is straightforward to show that for all states with maximally mixed marginal systems, every state whose entanglement is witnessed by $S_{\alpha}(A|B)<0$ must have its entanglement witnessed with any entropy of higher order $\alpha'>\alpha$. This comes from the fact that the Renyi entropy of order $\alpha$ is a monotonically decreasing function of $\alpha$.

As a particularly striking example of how sensitive these higher-order entropies can be, we consider the case of the $N=D^{2}$-dimensional Werner state, which is a mixture of the Bell state $|\Phi\rangle\langle\Phi|$ and the maximally mixed state:
\begin{align}
\rho_{AB}^{(Werner)}&=p |\Phi\rangle\langle\Phi| + (1-p)\frac{\mathbf{I}}{D^2},\\
&|\Phi\rangle\equiv\frac{1}{\sqrt{D}}\sum_{i=1}^{D}|i\rangle|i\rangle.
\end{align}
The probability eigenvalue vectors for the Werner state are:
\begin{align}
\vec{\lambda}(AB)&=\Big(p+\frac{1-p}{D^{2}},\frac{1-p}{D^{2}},...,\frac{1-p}{D^{2}}\Big),\\
\vec{\lambda}(A)&=\vec{\lambda}(B)=\Big(\frac{1}{D},...,\frac{1}{D}\Big).
\end{align}

The entanglement of the Werner state is witnessed whenever $S_{\alpha}(A|B)<0$. For constant $p$, $S_{\alpha}(A|B)$ decreases as $\alpha$ increases; and for constant $\alpha$, $S_{\alpha}(A|B)$ decreases as $p$ increases. To keep the value of $S_{\alpha}(A|B)$ constant at increasing $\alpha$, there must also be a corresponding decrease in $p$. The threshold Bell state fraction $p$ for which $S_{\alpha}(A|B)=0$ must also decrease as $\alpha$ increases. See plots in Fig.~2 for example. 

Clearly for Werner states, higher-order Renyi entropies make for more sensitive witnesses of entanglement than lower order. Indeed, if one uses $S_{1}(A|B)$, one finds that the threshold value of $p$, ($p_{c}$), does not scale favorably at high dimension. Instead, $p_{c}$ asymptotically approaches $1/2$ as $N\rightarrow\infty$. On the other hand, using $S_{2}(A|B)$ scales more favorably, and has an analytic value of $p_{c}=1/\sqrt{D+1}$ (where $N=D^{2}$), decreasing toward zero for large dimension. Going beyond second order, using $S_{\infty}(A|B)$ scales better still, with an analytic value of $p_{c}=1/(D+1)$, a quadratic improvement over the collision entropy. Indeed, it was shown in \cite{PhysRevA.59.4206} that for $D\otimes D$ Werner states, $p_{c}=1/(D+1)$ is the necessary and sufficient critical value distinguishing separable states from entangled ones. Even here, the favorability of the scaling is understated. Recall that the $127$-qubit state has dimension of $2^{127}\approx 1.7\times10^{38}$, and a Werner state of such a dimension can still have its entanglement witnessed by comparing purities for any Bell state fraction greater than $7.67\times10^{-20}$.

\begin{figure}[t]
\includegraphics[width=0.95\columnwidth]{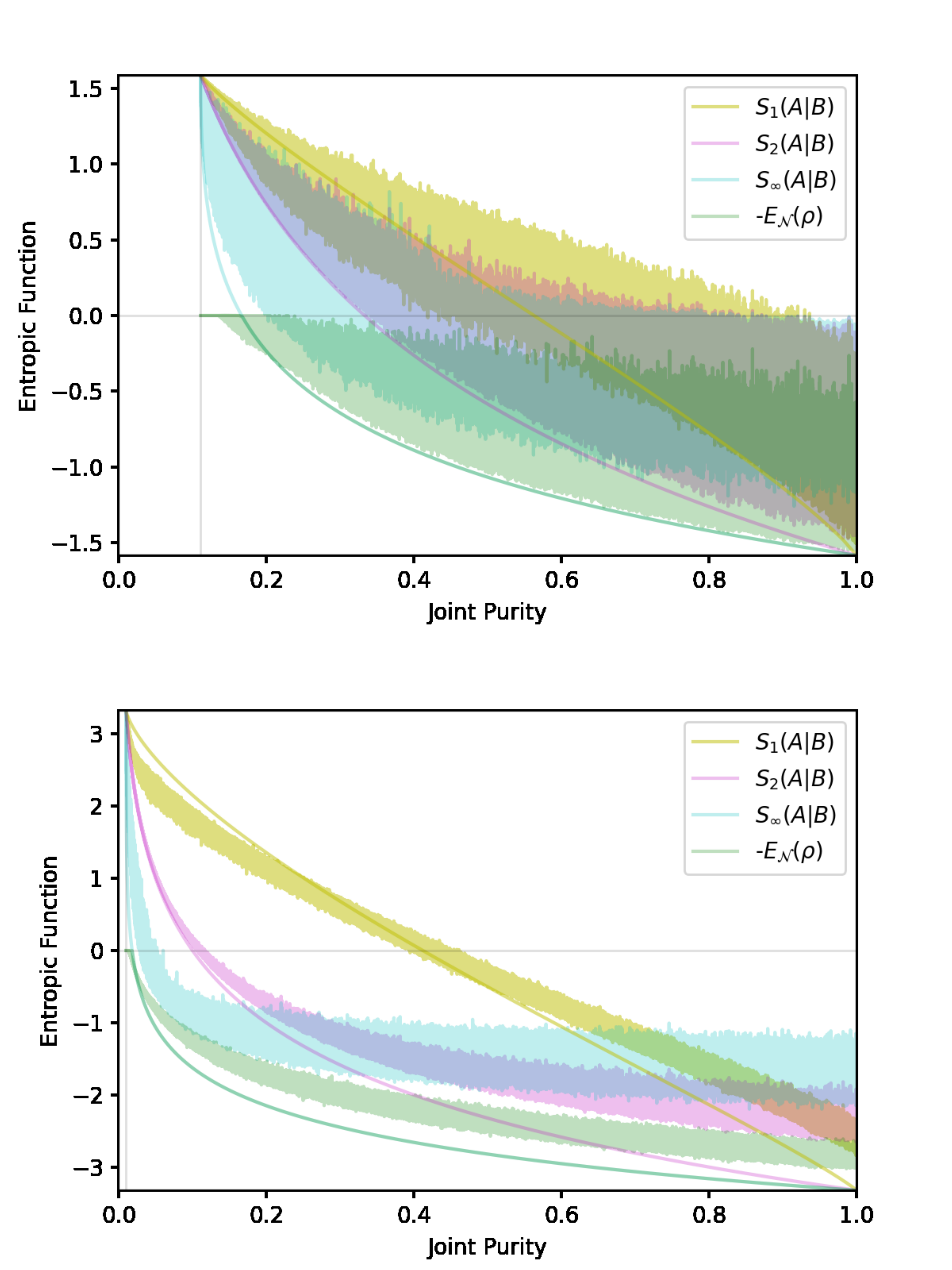}
\caption{Plots showing different conditional entropy functions (and the negativity) for uniform-purity ensembles as a function of joint purity. (Top) Case of $3\otimes 3$ systems. (Bottom) Case of $10\otimes 10$ systems. Narrow curves of the same color plot the corresponding function of the Werner state whose purity is varied by changing the mixing parameter $p$. Note that the Werner state curves for negative the log negativity $-E_{\mathcal{N}}(\hat{\rho})$ coincides with the conditional min entropy $S_{\infty}(A|B)$ where entanglement is witnessed.}
\end{figure}

To examine the more general case of success in entanglement witnessing, we have used the data from the uniform purity ensembles for $3\otimes 3$ and $10\otimes 10$ systems, and plotted the different conditional entropy functions as well as the (logarithmic) negativity $E_{\mathcal{N}}(\hat{\rho})$ as a function of joint state purity in Fig.~2. Although there is a substantial amount of noise at low dimension, we can clearly see as in the Werner state case, that the range of purities at which entanglement can be witnessed expands when using higher-order entropy.

\section{Upper and Lower bounds to von Neumann entropy given constant state purity}\label{bounds}
While comparing joint and marginal purities seems to be more effective at witnessing entanglement than comparing von Neumann entropies, it is von Neumann entropies that have a larger utility in various applications in quantum information science. To this end, it is useful to point out straightforward upper and lower bounds to the von Neumann entropy for a given state purity.

\begin{figure}[t]
\includegraphics[width=0.9\columnwidth]{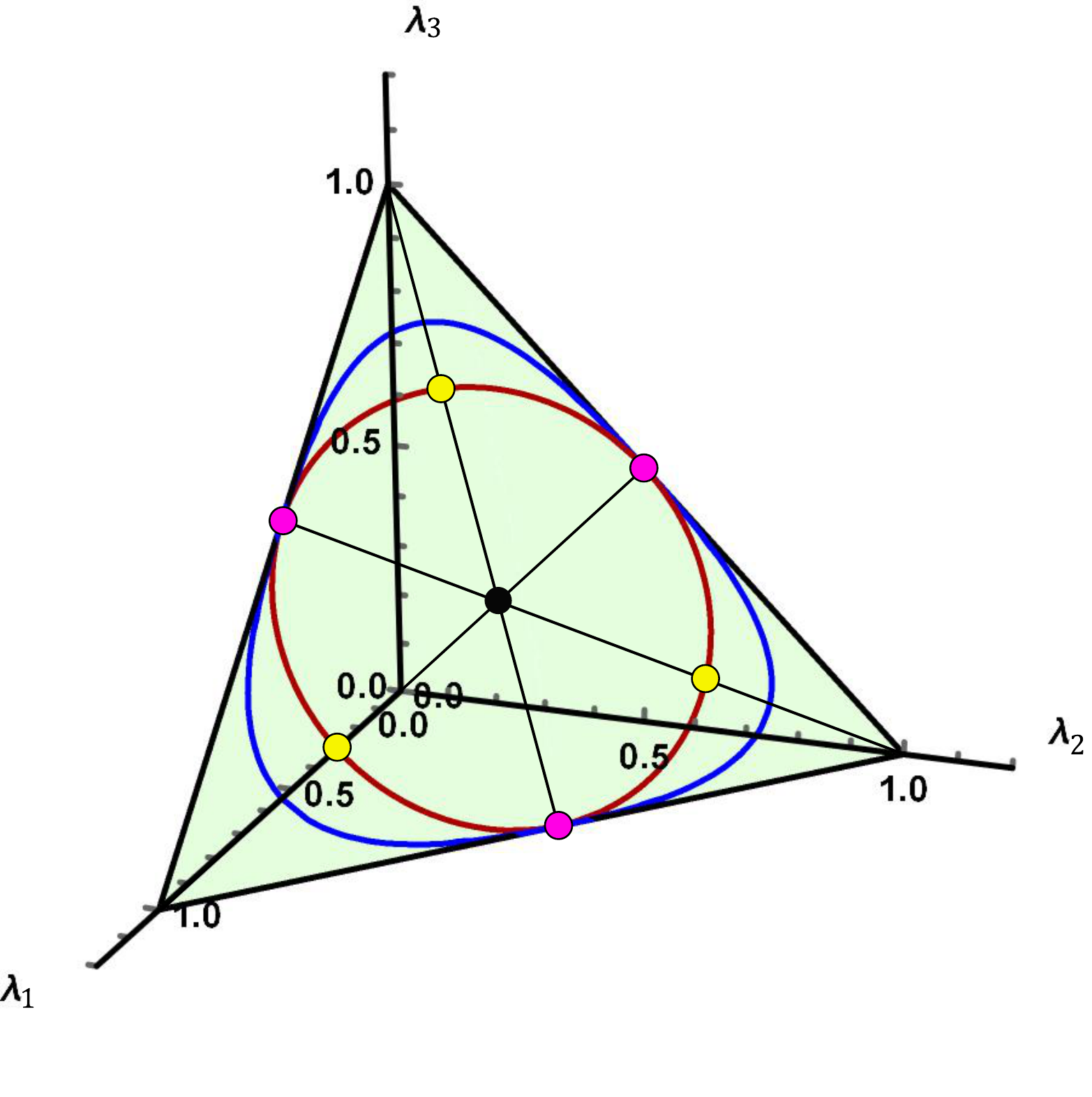}
\caption{Plot of probability simplex of three outcomes $(\lambda_{1}+\lambda_{2}+\lambda_{3}=1)$ with contours of constant purity (red circle) and constant von Neumann entropy (blue triangular loop), where $S_{1}=S_{2}=1$ bit. As purity and entropy decrease, the contours shrink, converging to the maximally mixed state at the centroid of the simplex (black dot). For a constant purity, the maximum entropy distributions will lie at the intersections between the purity contour and the medial line segments going from the centroid to the vertices of the simplex (yellow circles). For this dimension, the minimum entropy distributions lie at the opposite intersections (magenta circles).}
\end{figure}

In \cite{berry2003bounds}, Berry and Sanders provide these bounds, which we now discuss here and make a note in Appendix~\ref{pmin}. The maximum entropy distribution (given in equation \eqref{pmax} as $\vec{p}_{max}$) for constant purity $\mathcal{P}$ is uniform except for one outcome. In the probability simplex  of $N$ outcomes (see Fig.~3 for example), points on line segments going from the maximally mixed state to a vertex are the only distributions of this class, being equal in all coordinates except for one (not counting the maximally mixed state). Where the manifold of constant purity probability vectors forms an $N-2$ dimensional sphere centered on the maximally mixed state, the intersection of this manifold with these line segments gives the probability distributions of maximum entropy:
\begin{equation}\label{pmax}
\vec{p}_{max}=(p_{0},...,p_{0},1-(N-1)p_{0}).
\end{equation}
Given the form of $\vec{p}_{max}$, the purity $\mathcal{P}$ is readily expressed in terms of $p_{0}$:
\begin{equation}
\mathcal{P}=(N-1)p_{0}^{2} +(1-(N-1)p_{0})^2.
\end{equation}
Of the two values of $p_{0}$ satisfying this constraint, the value of $p_{0}$ less than $1-(N-1)p_{0}$ is the value producing the maximum entropy distribution.

The \emph{minimum} entropy probability distribution (given in equation \eqref{pmindist}) for constant purity $\mathcal{P}$ is a discrete top-hat distribution appended with a final nonzero probability and all other probabilities zero. If we define $\kappa$ as $1/\mathcal{P}$ rounded down to the nearest integer (see Appendix~\ref{pmin} for discussion), then the minimum entropy probability distribution will have $\kappa$ outcomes of equal nonzero probability, one outcome with lesser probability, and all other outcomes of probability zero:
\begin{equation}\label{pmindist}
\vec{p}_{min}=(p_{0},...,p_{0},1- \kappa p_{0},0,0,...,0).
\end{equation}
Except for the case of three dimensions $(N=3)$, $\vec{p}_{min}$ and $\vec{p}_{max}$ do not both fall on the same lines passing through the maximally mixed state (See Fig.~3 for a diagram of the 3-dimensional case). In general, there will be more distributions of minimum von Neumann entropy for a given state purity than maximum entropy distributions because all permutations of $\vec{p}$ have the same entropy, and there are never fewer permutations of $\vec{p}_{min}$ than of $\vec{p}_{max}$.

\section{Discussion: \\Correlations vs Negativity \\in witnessing entanglement}
In this work, we have shown that all methods of witnessing entanglement between two parties $A$ and $B$ by comparing the mixedness of the joint state to that of the marginals are subsumed by the negativity of the partial transpose. Given the close relationship between entanglement and correlation, we quickly discuss how many, but not all correlation-based witnesses are also subsumed by the negativity.

For two parties $A$ and $B$ sharing a pure quantum state between them $|\psi\rangle_{AB}$, all correlations are identifiable as entanglement, and the strength of those correlations corresponds to the amount of entanglement present. This relationship between correlation and entanglement is preserved for mixed states up to the amount of mixing present. For a pair of observables $\hat{X}_{A}$ and $\hat{X}_{B}$ of a joint quantum system with density operator $\hat{\rho}_{AB}$ whose correlations are quantified by the mutual information $H(X_{A}:X_{B})$, the relationship is given by the relation:
\begin{equation}\label{EntBoundCorrRel1}
H(X_{A}:X_{B}) \leq E_{F}(AB) + S(AB),
\end{equation}
where $E_{F}(AB)$ is the entanglement of formation of $\hat{\rho}_{AB}$ \cite{WootersEntForm1998}, and $S(AB)$ is the von Neumann entropy of $\hat{\rho}_{AB}$. See Appendix for proof. This relation shows that up to the mixedness of the joint state, there can be no correlations without entanglement. Importantly, this means that for nearly pure states with joint quantum entropy $S(AB)$ near zero, nearly all correlations can be identified as entanglement. However, this relation is based on negative values of the quantum conditional entropy $S(A|B)$ being lower limits to entanglement measures such as the entanglement of formation $E_{F}$. Because of this, many classes of correlation-based entanglement witnesses including many EPR-steering inequalities \cite{Berta2010, Walborn2011,Schneeloch2013, schneeloch2018quantifying} are subsumed by the negativity of the partial transpose, but not all of them.

In \cite{vertesi2014disproving}, there are Bell inequalities that can witness the entanglement in bound-entangled states with a positive partial transpose. Where Bell inequalities fundamentally identify entanglement through correlations, we see that the relationship between entanglement and correlations is more subtle than relative mixedness can describe.

\section{Conclusion: Merits of different entanglement witnesses}\label{conclusion}
In our investigations, we examined how well comparing the mixedness of a joint quantum state to the mixedness of its subsystems witnesses entanglement. While the von Neumann entropy is a popular measure of mixedness, we find that even comparing the joint and marginal purities (i.e., $\text{Tr}[\hat{\rho}^{2}]$) witnesses entanglement in more quantum states than when using the von Neumann entropy. This is promising, as there exist direct measurements of $\text{Tr}[\hat{\rho}^{2}]$ by interfering two copies of a quantum state \cite{PhysRevLett.88.217901,PhysRevLett.95.240407}, so that full state tomography is unnecessary.  When full state tomography \emph{is} possible, we proved that computing the negativity of the partial transpose is more sensitive at witnessing entanglement than comparing with any measure of mixedness.

\begin{acknowledgments}
We gratefully acknowledge support from the Air Force Office of Scientific Research LRIR 18RICOR028 as well as insightful discussions with Dr. Michael Fanto, Dr. Ashley Prater-Bennette, and Dr. Richard Birrittella. 

The views expressed are those of the authors and do not reflect the official guidance or position of the United States Government, the Department of Defense or of the United States Air Force. The appearance of external hyperlinks does not constitute endorsement by the United States Department of Defense (DoD) of the linked websites, or of the information, products, or services contained therein. The DoD does not exercise any editorial, security, or other control over the information you may find at these locations.
\end{acknowledgments}

\appendix
\section{Note on the expression of $\vec{p}_{min}$.}\label{pmin}
In \cite{berry2003bounds}, Berry and Sanders give the value $\kappa$ in equation \eqref{pmindist} as the value $1/p_{0}$ rounded down to the nearest integer instead of $1/\mathcal{P}$ rounded down to the nearest integer. In this Appendix, we show that both these expressions are equivalent.

Using the Berry-Sanders definition for $\kappa$, it follows that
\begin{equation}
\frac{1}{p_{0}}-1\leq\kappa\leq \frac{1}{p_{0}}
\end{equation}
Next, we know that $\vec{p}_{min}$ has between $\kappa$ and $\kappa+1$ outcomes, so it should be straightforward to prove that the purity $\mathcal{P}$ is bounded by:
\begin{equation}\label{Purbound}
\frac{1}{\kappa} \geq\mathcal{P}\geq \frac{1}{\kappa+1}
\end{equation}
To prove this, we note that $p_{0}$ is also bounded between $1/(\kappa+1)$ and $1/\kappa$, and that within this range the purity $\mathcal{P}$ increases monotonically with $p_{0}$:
\begin{align}
\mathcal{P}&=\kappa p_{0}^{2} + (1- \kappa p_{0})^2\\
\frac{\partial \mathcal{P}}{\partial p_{0}} &= 2\kappa (p_{0}(\kappa+1)-1)\\
\frac{\partial \mathcal{P}}{\partial p_{0}}&\geq 0\qquad\text{where}\qquad p_{0}\geq\frac{1}{\kappa+1}
\end{align}
From this, we can say that where $1/(\kappa+1)$ and $1/\kappa$ are valid values for the purity $\mathcal{P}$, $\mathcal{P}$ must increase monotonically as $p_{0}$ does from $1/(\kappa+1)$ to $1/\kappa$, proving equation \eqref{Purbound}.

Next, we can algebraically rearrange the inequalities in equation \eqref{Purbound} into the relation:
\begin{equation}
\frac{1}{\mathcal{P}}-1 \leq\kappa\leq \frac{1}{\mathcal{P}}
\end{equation}
As $\kappa$ is defined to be an integer, we can conclude that while $\kappa$ was originally defined as $1/p_{0}$ rounded down to the nearest integer, it is equivalent to the value of $1/\mathcal{P}$, rounded down to the nearest integer.

\section{The probability of ``Nearly pure'' and ``Nearly maximally mixed'' states}\label{nearlypure}
In Section~\ref{eigenspectrum}, we introduce the concept of a ``nearly pure'' probability vector as one in which the maximum component is greater than or equal to $1/2$. Geometrically, we can look at a vertex of the probability simplex, and select the halfway points on all lines connecting this vertex to other vertices. The convex hull of the vertex and this set of midway points is a simplex identical in shape to the probability simplex, but scaled down by a factor of 1/2 in every dimension. Within this sub-simplex there are only nearly pure states because its proximity to the vertex of the probability simplex requires that the maximum probability component of any point in this sub simplex be at least $1/2$.

The ratio of the volume of the sub-simplex to the total probability simplex is $(1/2^{N-1})$ where $N$ is the total dimension of the probability space (and the number of outcomes of the probability distribution). Moreover, there are only $N$ of these sub-simplices containing the nearly pure states because there are only $N$ vertices of the probability simplex. Putting these two facts together, the fraction of states on the probability simplex that are nearly pure $P_{\approx P}$ is: 
\begin{equation}
P_{\approx P}=\frac{N}{2^{N-1}}.
\end{equation}
For three dimensions (see Fig.~3), this fraction is $3/4$, but it decreases exponentially with dimension $N$.

To find the probability of the nearly maximally mixed states, we point out that the volume $V_{N}$ of the $N$-dimensional probability simplex is:
\begin{equation}
V_{N}^{(simplex)}=\frac{\sqrt{N}}{(N-1)!}.
\end{equation}
The set of nearly maximally mixed states is the set of states whose purity $\mathcal{P}$ is between $1/N$ and $1/(N-1)$. The volume in the simplex that these states occupy is of a uniform $N$-dimensional sphere, centered on the maximally mixed state, and with a radius $r=\sqrt{\frac{1}{N(N-1)}}$:
\begin{equation}
V_{N}^{(sphere)}=\frac{\pi^{\frac{N}{2}}}{\Gamma(\frac{N}{2}+1)}\left(r\right)^{N}
\end{equation}
The ratio of these two volumes gives the probability of a nearly maximally mixed state $P_{\approx MM}$ under uniform sampling to be:
\begin{equation}
P_{\approx MM}=\frac{1}{\sqrt{N}}\left(\frac{\pi}{N(N-1)}\right)^{N/2}\frac{\Gamma(N)}{\Gamma(N/2 +1)}
\end{equation}
Where $\Gamma(N)\leq (N/2+1)^{(N/2-1)}\Gamma(N/2+1)$, and where in general $\Gamma(N/2+1) \ll \Gamma(N)$, we have:
\begin{equation}
P_{\approx MM}\ll\frac{1}{(N/2)\sqrt{N}}\left(\frac{\pi}{N}\right)^{N/2}
\end{equation}
In short, we see that the probability of nearly maximally mixed states decreases even faster than exponentially under uniform dampling.

\section{The Law of Large Numbers \\and na\"{i}ve sampling of probability vectors}\label{naive}
In Section \ref{eigenspectrum}, we argue that a na\"{i}ve sampling of a probability vector $\vec{p}$ of dimension $N$ in which $N$ random numbers are drawn from a probability distribution on the real numbers between zero and unity produces vectors overwhelmingly weighted toward the maximally mixed state. In particular, we can prove that where $H(\vec{p})$ is the Shannon entropy of a probability vector $\vec{p}$ obtained from this na\"{i}ve sampling process, that:
\begin{equation}
H(\vec{p})\rightarrow \log(N) + const
\end{equation}
where $\log(N)$ is the entropy of a maximally mixed probability vector of dimension $N$.

The law of large numbers (in particular, Borel's law of large numbers \cite{BorelProof1984}) states that as the length of a sequence of independent and identically distributed random variables grows, the relative frequencies of each outcome of the random variable converge to their respective probabilities. If the probability of outcome $j$ of random variable $X$ is $p$, then for a sufficiently large number of trials, the fraction of trials for which the outcome of $X$ was $j$ will also converge to $p$.

Consider generating a probability vector $\vec{p}$ by the na\"{i}ve sampling described at the beginning of this section. In each trial, we generate $N$ random values according to the same fixed probability distribution for each value, and normalize them to randomly generate the $N$ components of $\vec{p}$. Let the function $f(k)$ where $k$ goes from $1$ to $N$ define the $k$th smallest generated probability component in $\vec{p}$ from this random sample. Next, we define the rescaled function $\tilde{f}(x)=N\times f(Nx)$ so that $x$ goes from $1/N$ to one as $k$ goes from one to $N$.

As the dimension $N$ grows, the law of large numbers proves that  $\tilde{f}(x)$ converges to some fixed probability density on the interval $[0,1]$, so that $f(k)$ forms a scaled version of $\tilde{f}(x)$ by factor $N$. This implies that the entropy of the na\"{i}vely generated probability vector $\vec{p}$ approaches $\log(N)$ plus a constant equal to the continuous entropy of $\tilde{f}(x)$: 
\begin{equation}
H(\vec{p})\rightarrow \log(N) + h(\tilde{f}(x))
\end{equation}
Here, we point out that $h(\tilde{f}(x))$ is the continuous Shannon entropy of probability density $\tilde{f}(x)$ \cite{Cover2006}; it is independent of dimension $N$, and becomes insignificant in the limit of large  $N$. Where $\log(N)$ is the entropy of a maximally mixed probability vector of dimension $N$, it follows that probability vectors generated in this na\"{i}ve fashion will be substantially biased toward maximally mixed distributions.

\begin{figure}[t]
\centerline{\includegraphics[width=0.9\columnwidth]{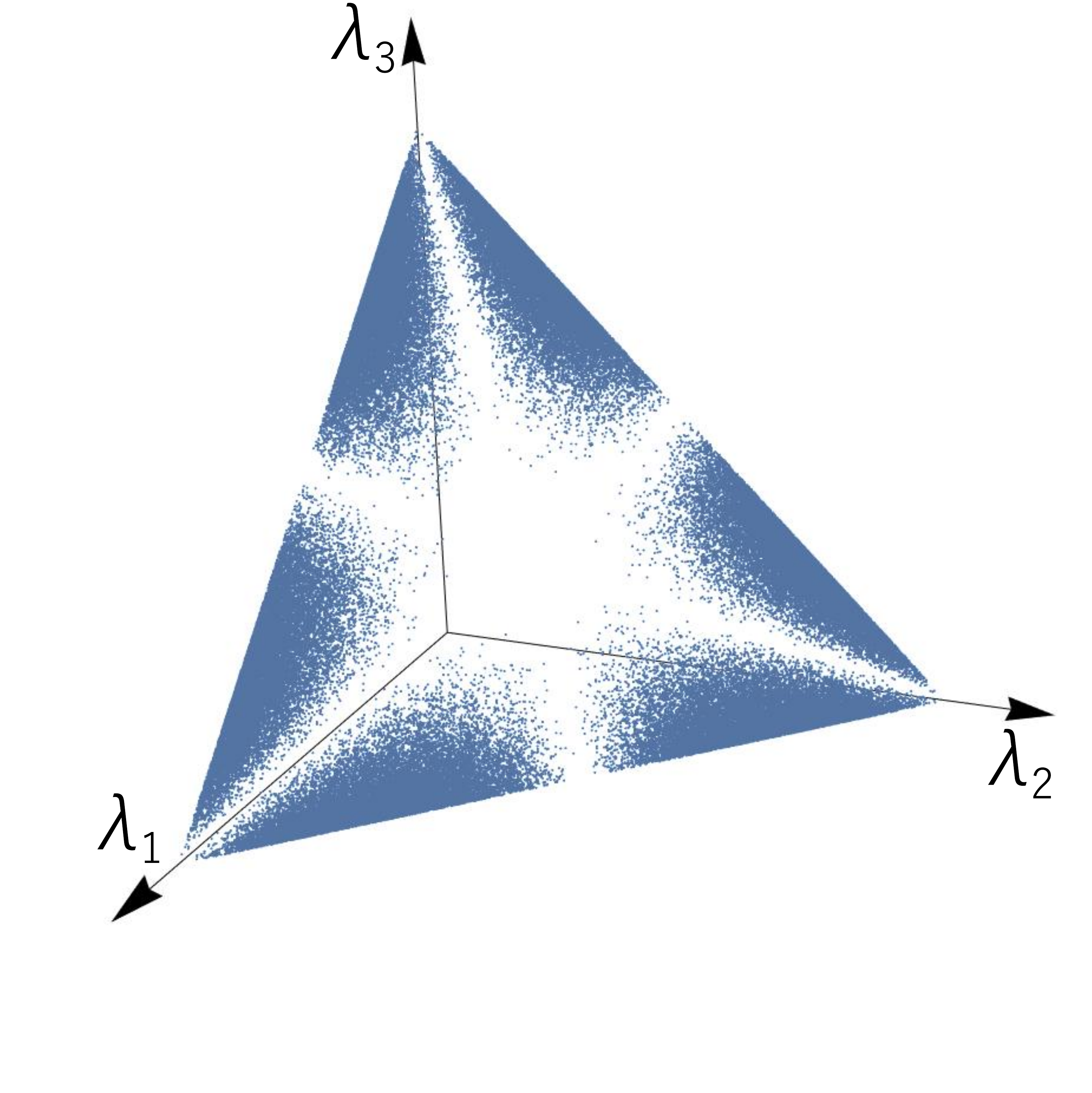}}
\caption{Scatterplot of marginal eigenvalue vectors $\vec{\lambda}=(\lambda_{1},\lambda_{2},\lambda_{3})$ of the partial trace of uniformly sampled joint $3\otimes 3$ pure states.}
\end{figure}

\section{Rarity of highly entangled states even when uniformly sampling pure states}\label{MargEnt}
The algorithm for generating a uniform distribution of density matrices where: one takes a random eigenvalue vector uniform on the probability simplex, and then transforms the diagonal matrix with a random unitary matrix uniform on the Haar measure, is well-justified. However, when one interprets this uniform distribution of density matrices as a uniform distribution of $N=D\otimes D$ joint density matrices, the resulting distribution of marginal density matrices is far from uniform.

As an example, we consider the uniform distribution of $D\otimes D$ pure states generated from Haar-random unitaries acting on a single pure state. In \cite{ZyczkowskiBook}, this distribution of marginal eigenvalues is actually derived explicitly. The distribution of the eigenvalues of the $D$-dimensional subsystem taken from the partial trace of uniformly sampled $D\otimes D$-dimensional pure states is given by:
\begin{align}
\rho(\vec{\lambda},D)&=\frac{\Gamma(D^{2})}{\prod_{j=0}^{D-1}(\Gamma(D-j)\Gamma(D+1-j))}\nn\\
&\cdot \delta(1-\sum_{i}\lambda_{i})\prod_{i<j}^{D}(\lambda_{i}-\lambda_{j})^{2}.
\end{align}
where $\Gamma(x)$ is Euler's gamma function, and $\delta(x)$ is the Dirac delta function, employed here to enforce the constraint that the eigenvalues sum to unity.

In particular, we note that anywhere in the (marginal) probability simplex where two or more eigenvalues approach equality, the probability of generating such an eigenvalue vector approaches zero. See Fig.~4 for $3\otimes 3$ example. Where the maximally entangled states occur as the marginal eigenvalues approach $1/D$ (equaling each other), the probability of generating highly entangled states becomes vanishingly small even when accounting for the fraction of the volume of the simplex where such states would reside.

\section{Proof of entanglement-correlation relation}
In Section VI, we illustrated the relationship between correlation and entanglement through the relation:
\begin{equation} \label{ECorRel}
H(X_{A}:X_{B}) \leq E_{F}(AB) + S(AB),
\end{equation}
This relationship is straightforward to show because the classical correlations between two parties $A$ and $B$ is bounded from above by the marginal quantum entropy demonstrated in \cite{Hall2006}:
\begin{equation}\label{corrReal}
H(X_{A}:X_{B}) \leq \min\{S(A),S(B)\},
\end{equation}
To prove this, we follow the basic logic laid out in \cite{Hall2006}. First, we note that the classical mutual information $H(X_{A}:X_{B})$ is the quantum mutual information of $AB$ after observables $\hat{X}_{A}$ and $\hat{X}_{B}$ have been measured. Because quantum mutual information is a form of relative entropy, we can use the monotonicity of relative entropy to say that the quantum mutual information decreases upon measurement, and in particular, that:
\begin{equation}
H(X_{A}:X_{B})\leq H(X_{A}:B)
\end{equation}
where the right-hand side is the quantum mutual information after only $\hat{X}_{A}$ has been measured.

After $\hat{X}_{A}$ has been measured, the joint state $\hat{\rho}_{AB}$ has a form known as classical-quantum:
\begin{equation}
\hat{\rho}_{X_{A}B}=\sum_{i} P(X_{Ai})|X_{Ai}\rangle\langle X_{Ai}|\otimes \hat{\rho}_{Bi}
\end{equation}
The mutual information of this kind of state \cite{wilde2017quantum}: is given by:
\begin{equation}
H(X_{A}:B)=S(B) - \sum_{i} P(X_{Ai}) S(\hat{\rho}_{Bi})
\end{equation}
Since von Neumann entropy (without conditioning) is non negative, we obtain the bound:
\begin{equation}
H(X_{A}:B)\leq S(B)
\end{equation}
Together, this gives us the relation:
\begin{equation}
H(X_{A}:X_{B})\leq S(B).
\end{equation}
Since we can perform the same set of steps with system $B$ being measured instead, we see that $H(X_{A}:X_{B})$ cannot exceed either $S(A)$ or $S(B)$, thus proving the correlation relation \eqref{corrReal} demonstrated in \cite{Hall2006}. 

Next, we express $\min\{S(A),S(B)\}$ in terms of conditional and joint entropy:
\begin{align}
\min\{S&(A),S(B)\}= \min
\begin{Bmatrix}
S(AB)-S(B|A)\\
 ,S(AB) -S(A|B)
 \end{Bmatrix}\nn\\
&=\min\{-S(B|A), -S(A|B)\}+ S(AB) 
\end{align}
Where the quantum conditional entropy is concave, we have for any pure state decomposition:
\begin{align}
\min\{S&(A),S(B)\}\leq S(AB) +\nn\\
&+\min\left\{\sum_{i} p_{i}(-S_{i}(B|A)), \sum_{i} p_{i}(-S_{i}(A|B))\right\} \nn\\
&=\min\left\{\sum_{i} p_{i}S_{i}(A), \sum_{i} p_{i}S_{i}(B)\right\}+ S(AB) 
\end{align}
Where the same pure state decomposition is taken for both sums, we must have that $S_{i}(A)=S_{i}(B)$, and:
\begin{align}
\min\{S&(A),S(B)\}\leq S(AB) +\nn\\
&+\min\left\{\sum_{i} p_{i}(-S_{i}(B|A)), \sum_{i} p_{i}(-S_{i}(A|B))\right\} \nn\\
&=\min\left\{\sum_{i} p_{i}S_{i}(A)\right\}+ S(AB) 
\end{align}
Since this relation must be true for all pure state decompositions, we can choose the minimizing pure-state decomposition defining the entanglement of formation:
\begin{equation}
E_{F}(AB)=\min_{|\psi\rangle}\left\{\sum_{i} p_{i}S_{i}(A)\right\}
\end{equation}
which together with the previous inequality gives us the relation:
\begin{equation}
\min\{S(A),S(B)\} \leq E_{F}(AB) + S(AB).
\end{equation}
Incorporating this into the correlation relation \eqref{corrReal} gives us the entanglement-correlation relation:
\begin{equation}\label{thisrelation}
H(X_{A}:X_{B})\leq E_{F}(AB) + S(AB).
\end{equation}
\emph{Important note}: Because this relation \eqref{thisrelation} exists at all finite dimension with neither side explicitly dependent on dimension, it exists in the continuum limit \footnote{See \cite{schneeloch2018quantifying} for example of use of continuum limit}. Where the mutual information between a pair of continuous observables is given by $h(x_{A}:x_{B})$, the corresponding entanglement-correlation relation is given by:
\begin{equation}
h(x_{A}:x_{B})\leq E_{F}(AB) + S(AB).
\end{equation}

\bibliography{EPRbib17}
%
%
%
%
%


\end{document}